\documentclass[12pt]{article}
\usepackage{a4wide}
\usepackage{epsfig}
\usepackage{amsmath}

\newlength{\absize}
\catcode`@=11
\def\citer{\@ifnextchar [{\@tempswatrue\@citexr}{\@tempswafalse\@citexr[]}}

\def\@citexr[#1]#2{\if@filesw\immediate
  \write\@auxout{\string\citation{#2}}\fi
  \def\@citea{}\@cite{\@for\@citeb:=#2\do
    {\@citea\def\@citea{--\penalty\@m}\@ifundefined
       {b@\@citeb}{{\bf ?}\@warning
       {Citation `\@citeb' on page \thepage \space undefined}}%
\hbox{\csname b@\@citeb\endcsname}}}{#1}}
\catcode`@=12

\begin{document}
  \thispagestyle{empty}
  \pagestyle{empty}
  \renewcommand{\thefootnote}{\fnsymbol{footnote}}
\newpage\normalsize
    \pagestyle{plain}
    \setlength{\baselineskip}{4ex}\par
    \setcounter{footnote}{0}
    \renewcommand{\thefootnote}{\arabic{footnote}}
\newcommand{\preprint}[1]{%
  \begin{flushright}
    \setlength{\baselineskip}{3ex} #1
  \end{flushright}}
\renewcommand{\title}[1]{%
  \begin{center}
    \LARGE #1
  \end{center}\par}
\renewcommand{\author}[1]{%
  \vspace{2ex}
  {\Large
   \begin{center}
     \setlength{\baselineskip}{3ex} #1 \par
   \end{center}}}
\renewcommand{\thanks}[1]{\footnote{#1}}
\begin{flushright}
\end{flushright}
\vskip 0.5cm

\begin{center}
{\large \bf Fractional Angular Momentum in Noncommutative Space}
\end{center}
\vspace{1cm}
\begin{center}
Jian-zu Zhang$^{a,b,\#}$
\end{center}
\vspace{1cm}
\begin{center}
$^a$ Institute for Theoretical Physics, Box 316, East China
University of Science and Technology, Shanghai 200237, P. R. China \\
$^b$ Department of Physics, University of Kaiserslautern, PO Box
3049, D-67653 Kaiserslautern, Germany
\end{center}
\vspace{1cm}

\begin{abstract}

In noncommutative space to maintain Bose-Einstein statistics for
identical particles at the non-perturbation level described by
deformed annihilation-creation operators when the state vector
space of identical bosons is constructed by generalizing
one-particle quantum mechanics it is explored that the consistent
ansatz of commutation relations of phase space variables should
simultaneously include space-space noncommutativity and
momentum-momentum noncommutativity, and a new type of boson
commutation relations at the deformed level is obtained.
Consistent perturbation expansions of deformed
annihilation-creation operators are obtained. The influence of the
new boson commutation relations on dynamics is discussed. The
non-perturbation and perturbation property of the orbital angular
momentum of two-dimensional system are investigated. Its spectrum
possesses fractional eigen values and fractional intervals.

\end{abstract}

\begin{flushleft}
${^\#}$ E-mail address: jzzhang@ecust.edu.cn  \\
\hspace{3.2cm} jzzahng@physik.uni-kl.de

\end{flushleft}
\clearpage

Recently there has been a renewed interest in physics in
 noncommutative space \citer{CDS,SE}.
This is motivated by studies of the low energy effective theory of
D-brane with a nonzero NS-NS $B$ field background. The effects of
noncommutative space only appear near the string scale, thus to
test the space noncommutativity we need noncommutative quantum
field theories. But study at the level of quantum mechanics in
noncommutative space is meaningful for clarifying some possible
phenomenological consequences in solvable models. In literature
the perturbation aspects of noncommutative quantum mechanics(NCQM)
have been extensively studied \citer{CST,RS}. The perturbation
approach is based on the Weyl-Moyal correspondence \citer{AW,RS},
according to which the usual product of functions should be
replaced by the star-product.

In this paper we are interested in the non-perturbation
investigation which may explores some essentially new features of
NCQM. Because of the exponential differential factor in the
Weyl-Moyal product the non-perturbation treatment is difficulty. A
suitable example for the non-perturbation investigation is two
dimensional isotropic harmonic oscillator. This model is exactly
soluble, and explored fully, for example, in \cite{NP,HS,SS}. In
the following through the non-perturbation investigation of this
example we clarify the consistent condition for quantum theories
in noncommutative space in detail. The point is how to maintain
Bose-Einstein statistics at the non-perturbation level described
by deformed annihilation-creation operators  in noncommutative
space when the state vector space of identical bosons is
constructed by generalizing one-particle quantum mechanics. We
find that for this purpose the consistent ansatz of commutation
relations of phase space variables should simultaneously include
space-space noncommutativity and momentum-momentum
noncommutativity, and a new type of boson commutation relations at
the deformed level is obtained. To the linear terms of undeformed
annihilation-creation operators,
 perturbation expansions of deformed annihilation-creation operators
are obtained. The influence of the new deformed boson
commutation relations on dynamics is investigated.
As an example, we study
the angular momentum, and explore that its spectrum possesses
fractional eigen values and
fractional
intervals, specially there is a fractional zero-point angular
momentum. The perturbation treatment of
the harmonic oscillator is briefly presented.

In order to develop the NCQM formulation we
need to specify the phase space and the Hilbert space on which
operators act. The Hilbert space can consistently be taken to be
exactly the same as the Hilbert space of the corresponding
commutative system \citer{CST}.

As for the phase space we consider both space-space
noncommutativity (space-time noncommutativity is not considered)
and momentum-momentum noncommutativity. There are different types
of noncommutative theories, for example, see a review
paper \cite{DN}.

The former is inferred from the string theory \cite{AAS,CH99}. The
reasons of considering momentum-momentum noncommutativity are:
(i) To
maintain Bose-Einstein statistics for identical
particles at the level of the deformed
annihilation-creation operators when the state vector space of
identical bosons is constructed by generalizing one-particle
quantum mechanics;
(ii) To incorporate an additional background magnetic field
\cite{DN,NP}. The points (1) will become clear latter.

In the case of simultaneously space-space
 noncommutativity and momentum-momentum noncommutativity the
 consistent NCQM algebra is as follows (henceforth, $c$ and
 $\hbar$ are set to unit):
\begin{equation}
\label{Eq:xp} [\hat x_{i},\hat x_{j}]=i\xi^{-2}\Lambda_{NC}^{-2}
d\epsilon_{ij}, \qquad [\hat x_{i},\hat p_{j}]=i\delta_{ij},
\qquad [\hat p_{i},\hat p_{j}]=i\xi^{-2}\Lambda_{NC}^2
d^{\;\prime}\epsilon_{ij},\;(i,j=1,2)
\end{equation}
where $d$ and $d^{\;\prime}$ $\;$ are the constant,
frame-independent dimensionless parameters, $\epsilon_{ij}$ is a
antisymmetric unit tensor, $\epsilon_{12}=-\epsilon_{21}=1,$
$\epsilon_{11}=\epsilon_{22}=0.$ $\;$ The parameter $\Lambda_{NC}$
is the NC energy scale \cite{HPR} (We reset
$\theta_{ij}=\Lambda_{NC}^{-2}d\epsilon_{ij}$ where $\theta_{ij}$
is the noncommutative parameter extensively adopted in literature
for the case that only space-space are noncommuting ). The NC
effects will only become apparent as this scale is approached. In
Eq.~(\ref{Eq:xp}) $\xi$ is the scaling factor
$\xi=(1+dd^{\;\prime}/4)^{1/2}$. When $d^{\;\prime}=0,$ we have
$\xi=1,$ the NCQM algebra (\ref{Eq:xp}) reduces to the one which
is extensively discussed in literature for the case that only
space-space are noncommuting. Here and in the following $\hat F$
represents the operator in noncommutative space, and $F$ the
corresponding one in commutative space.

The NCQM algebra (\ref{Eq:xp}) changes the algebra of
creation-annihilation operators. There are
different ways to construct the creation-annihilation operators.
In order to explore noncommutative effects at the
non-perturbation level the
deformed creation-annihilation operators are
constructed which are related to the noncommutative variables
$\hat x_i,
\hat p_i.$

The Hamiltonian for two-dimensional isotropic harmonic
oscillator is (henceforth, summation convention is used)
\begin{equation}
\label{Eq:H} \hat H(\hat x,\hat p)=
\frac{1}{2\mu}\hat p_i\hat p_i
+\frac{1}{2}\mu\omega^2 \hat x_i\hat x_i.
\end{equation}
This system can be decomposed into two one-dimensional
oscillators. For the dimension $i$ the representations of the
deformed annihilation-creation operators $\hat a_i$, $\hat
a_i^\dagger$ $(i=1,2)$ are defined by
\begin{equation}
\label{Eq:aa+1} \hat a_i=\sqrt{\frac{\mu\omega}{2}}\left (\hat
x_i +\frac{i}{\mu\omega}\hat p_i\right), \quad
\hat
a_i^\dagger=\sqrt{\frac{\mu\omega}{2}}\left (\hat
x_i -\frac{i}{\mu\omega}\hat p_i\right).
\end{equation}
In order to maintain the physical meaning of $\hat a_i$ and $\hat
a_i^\dagger$ the relations among $(\hat a_i, \hat a_i^\dagger)$
and $(\hat x_i, \hat p_i)$ should keep the same formulation as the
ones in commutative space.

From Eq.~(\ref{Eq:aa+1})
and the NCQM algebra
(\ref{Eq:xp}) we obtain the commutator between the operators $\hat
a_{i}^\dagger$ and $\hat a_{j}^\dagger$:
$\left[\hat a_i^\dagger,\hat
a_j^\dagger\right]=\frac{i}{2}\xi^{-2}\mu\omega\left
(\Lambda_{NC}^{-2}d-\mu^{-2}\omega^{-2}
\Lambda_{NC}^2d^{\;\prime}\right)\epsilon_{ij}\equiv D_{ij}$.
If momentum-momentum is commuting, $d^{\;\prime}=0$,
for the case $i\ne j$ we would have
$D_{ij}\ne 0$ which would lead to that $\hat a_{i}^\dagger$ and
$\hat a_{j}^\dagger$ would not commute. When the state vector
space of identical bosons is constructed by generalizing
one-particle quantum mechanics, a natural requirement is that
Bose-Einstein statistics should be maintained at the
non-perturbation
level described by $\hat a_i^\dagger.$ If $D_{ij}$ were not zero,
when the operators $\hat a_1^\dagger\hat a_2^\dagger$ and $\hat
a_2^\dagger\hat a_1^\dagger$ were applied successively to a state,
say vacuum state $|0,0\rangle,$
\footnote{As in the case of commutative space the vacuum state $
|0,0\rangle$ in noncommutative space is defined as
\begin{equation*}
\hat a_i|0,0\rangle=0, \quad (i=1,2). \nonumber
\end{equation*}
}
would not produce the same physical state: $\quad$ $\hat
a_1^\dagger\hat a_2^\dagger |0,0\rangle-\hat a_2^\dagger\hat
a_1^\dagger|0,0\rangle=D_{ij}|0,0\rangle\ne 0.$ In order to
maintain Bose-Einstein statistics at the deformed level
the basic assumption is that
operators $\hat a_i^\dagger$ and $\hat a_j^\dagger$ are commuting,
that is, $D_{ij}\equiv 0.$ This requirement leads to a
consistency condition
\begin{equation}
\label{Eq:dd} d^{\;\prime}=\mu^2\omega^2\Lambda_{NC}^{-4} d.
\end{equation}
From Eqs.~(\ref{Eq:xp}), (\ref{Eq:aa+1})
and
(\ref{Eq:dd}) it follows that the commutation relations of $\hat
a_i$ and $\hat a_j^\dagger$ read
\begin{equation}
\label{Eq:[a,a+]1} \left[\hat a_1,\hat
a_1^\dagger\right]=\left[\hat a_2,\hat a_2^\dagger\right]=1
\quad \left[\hat a_i,\hat a_j\right]=\left[\hat
a_i^\dagger,\hat a_j^\dagger\right]=0,\;(i,j=1,2)
\end{equation}
\begin{equation}
\label{Eq:[a,a+]2} \left[\hat a_1,\hat a_2^\dagger\right]
=i\xi^{-2}\mu\omega\Lambda_{NC}^{-2} d.
\end{equation}
Eqs.~(\ref{Eq:[a,a+]1}) is the same commutation relations as the
ones in the commutative space. This confirms that for the same
degree of freedom $i$ the operators
 $\hat a_i,\hat a_i^\dagger$ are the correct deformed
annihilation-creation operators.
For the different degrees of freedom new deformed commutation
relations (\ref{Eq:[a,a+]2}) between $\hat a_i$ and
$\hat a_j^\dagger$ emerge.

We emphasize that Eq.~(\ref{Eq:[a,a+]2}) is consistent with {\it all}
principles of quantum mechanics and Bose-Einstein statistics.
\footnote{As a  check on the consistency , we apply
Eq.~(\ref{Eq:[a,a+]2}) to the vacuum state $ |0,0\rangle.$ Because
of $\hat a_i|0,0\rangle=0,$ the left hand side reads $(\hat
a_1\hat a_2^\dagger-\hat a_2^\dagger\hat a_1)|0,0\rangle=\hat
a_1\hat a_2^\dagger|0,0\rangle.$ In view of the noncommutativity
between $\hat a_1$ and $\hat a_2^\dagger,$ the term $\hat a_1\hat
a_2^\dagger|0,0\rangle\ne 0.$ The correct result can be obtained
by using the perturbation expansion (\ref{Eq:hat-a-a1}) of $\hat
a_i$ and $\hat a_{i}^\dagger:$ $\hat a_1\hat
a_2^\dagger|0,0\rangle=\xi^{-2}\left(a_1
a_2^\dagger+\frac{i}{2}\mu\omega\Lambda_{NC}^{-2}d\; a_1
a_1^\dagger+\frac{i}{2}\mu\omega\Lambda_{NC}^{-2}d\;a_2
a_2^\dagger-\frac{1}{4}\mu^2\omega^2\Lambda_{NC}^{-4}d^2\;a_2
a_1^\dagger\right)|0,0\rangle=i\xi^{-2}\mu\omega\Lambda_{NC}^{-2}d
|0,0\rangle$ which equals to the right
hand side.
 (In the last step the undeformed relations
 $a_i a_j^\dagger-a_j^\dagger a_i=\delta_{ij},$ $a_i|0,0\rangle=0$
and $a_i a_j^\dagger|0,0\rangle=\delta_{ij}|0,0\rangle$ are used.)}

If momentum-momentum is
commuting, $d^{\;\prime}= 0,$
which shows
that we could
not obtain $D_{ij}=0$.
Thus it is clear that in order to
maintain Bose-Einstein statistics for identical bosons at the
deformed
level of $\hat a_i$ and $\hat a_i^\dagger$ we should consider both
space-space noncommutativity and momentum-momentum
noncommutativity.

Now we consider perturbation expansions of $(\hat x_i, \hat p_j)$
and $(\hat a_i, \hat a_j^\dagger).$ The NCQM algebra (\ref{Eq:xp})
has different possible perturbation realizations \cite{NP}. To the
linear terms of phase space variables in commutative space, a
consistency ansatz of the perturbation expansions of $\hat x_{i}$
and $\hat p_{i}$ is
\begin{equation}
\label{Eq:hat-x-x} \hat
x_{i}=\xi^{-1}(x_{i}-\frac{1}{2}\Lambda_{NC}^{-2}d
\epsilon_{ij}p_{j}),
\quad
\hat
p_{i}=\xi^{-1}(p_{i}+\frac{1}{2}\mu^2\omega^2\Lambda_{NC}^{-2}
d\epsilon_{ij}x_{j}).
\end{equation}
where $x_{i}$ and $p_{i}$ are the coordinate and momentum in
commutative space and satisfy the commutation relations $\quad
[x_{i},x_{j}]=[p_{i},p_{j}]=0, \quad [x_{i},p_{j}]=i\delta_{ij}.$
In commutative space the undeformed annihilation-creation
operators $(a_i, a_i^\dagger)$ are related to the variables $(x_i,
p_i)$ by
\begin{equation}
\label{Eq:aa+3} x_i=\sqrt{\frac{1}{2\mu\omega}}\left (a_i
+a_i^\dagger\right), \quad
p_i=\frac{1}{i}\sqrt{\frac{\mu\omega}{2}}\left
(a_i -a_i^\dagger\right),
\end{equation}
where $a_i$ and $a_i^\dagger$ satisfy commutation relations
$[a_{i},a_{j}]=[a_i^\dagger,a_j^\dagger]=0, \quad
[a_{i},a^{\dagger}_{j}]=i\delta_{ij}.$
Inserting Eqs.~(\ref{Eq:hat-x-x}) and (\ref{Eq:aa+3}) into
Eq.~(\ref{Eq:aa+1}),
to the linear
terms of $a_{i}$ and $a_{i}^\dagger$, we obtain the perturbation
expansions of $\hat a_i$ and $\hat a_{i}^\dagger$ as follows
\begin{equation} 
\label{Eq:hat-a-a1} \hat
a_{i}=\xi^{-1}\left(a_{i}+\frac{i}{2}\mu\omega\Lambda_{NC}^{-2}
d\epsilon_{ij}a_j\right),\quad
\hat a_{i}^\dagger=\xi^{-1}\left(a_{i}^\dagger
-\frac{i}{2}\mu\omega\Lambda_{NC}^{-2}
d\epsilon_{ij}a_j\right).
\end{equation}

It is easy to check that Eqs.~(\ref{Eq:xp}), (\ref{Eq:aa+1})-
(\ref{Eq:hat-a-a1}) are
consistent each other.

Comparing to the case in commutative space the deformed
commutation relations (\ref{Eq:[a,a+]2}) change dynamical
properties of quantum theories in noncommutative space. As an
example, in the following  we investigate the influence of
Eq.~(\ref{Eq:[a,a+]2}) on the angular momentum. In two-dimensions
the orbital angular momentum is defined as an exterior product,
\footnote{There are different ways to define the angular momentum
in noncommutative space. In \cite{NP} comparing to the case in
commutative space, the angular momentum acquires $d-$ and
$d^{\;\prime}-$dependent scalar terms $\hat x_i \hat x_i$ and
$\hat p_i \hat p_i$. Because of the scalar terms have nothing to
do with the angular momentum, so in this paper we prefer to keep
the same definition of $\hat L$ as the one in commutative space.
The equation (\ref{Eq:[a,a+]2}) modifies the commutation relations
between $\hat L$ and $\hat x_i,$ $\hat p_i$. From the NCQM algebra
(\ref{Eq:xp}) we obtain
\begin{equation*}
 [\hat L,\hat x_i]=i\epsilon_{ij}\hat
x_j+i\xi^{-2}\Lambda_{NC}^{-2} d\;\hat p_i, \quad
 [\hat L,\hat
p_i]=i\epsilon_{ij}\hat p_j-i\xi^{-2}\mu^2\omega^2
\Lambda_{NC}^{-2} d\;\hat x_i.\nonumber
\end{equation*}
Comparing to the commutative case, the above commutation relations
acquires $d-$ and $d^{\;\prime}-$dependent terms which represent
effects in noncommutative space.}
\begin{equation}
\label{Eq:L0} \hat L=\epsilon_{ij}\hat x_i\hat p_j,
\end{equation}
Though $\hat L$ defined in Eq.~(\ref{Eq:L0}) has the same
formulation as the one in commutative space, because of the
commutation relations (\ref{Eq:[a,a+]2}) new features appear in
its spectrum.

 Using Eq.~(\ref{Eq:aa+1}) 
 we rewrite $\hat L$ as
\begin{equation}
\label{Eq:L1} \hat L=-i\left(\hat a_1^\dagger\hat a_2 -\hat
a_2^\dagger\hat a_1\right)-L_0,\quad L_0=\xi^{-2}
\mu\omega\Lambda_{NC}^{-2} d.
\end{equation}
Where the zero-point angular momentum
$L_0=\langle 0,0|\hat L|0,0\rangle$ originates from the
deformed commutation relations (\ref{Eq:[a,a+]2}).

In order to clarify the origin of the zero point angular momentum
$L_0$ from another point of view, in the following we investigate
the perturbation expansion of $\hat L.$ Starting from
Eq.~(\ref{Eq:L0}), using Eqs.~(\ref{Eq:hat-x-x})
we obtain
\begin{equation}
\label{Eq:L2} \hat
L=\epsilon_{ij}x_ip_j-\xi^{-2}\mu\Lambda^{-2}_{NC}
d\left(\frac{1}{2\mu}p_i p_i +\frac{1}{2}\mu\omega^2 x_i
x_i\right).
\end{equation}
Eq.~(\ref{Eq:L2}) is exactly solvable. We change the variables
$x_i$ and $p_i$ to variables $X_{\alpha}$ and $P_{\alpha}$ (here
and in the following $\alpha, \beta=a,b$) as follows
\cite{Bax,JZZ},
\begin{eqnarray}
\label{Eq:Xa} X_a&=&\sqrt{\frac{\mu\omega}{2\Omega_a}}\;x_1
-\sqrt{\frac{1}{2\mu\omega\Omega_a}}\;p_2, \quad
X_b=\sqrt{\frac{\mu\omega}{2\Omega_b}}\;x_1
+\sqrt{\frac{1}{2\mu\omega\Omega_b}}\;p_2, \nonumber\\
P_a&=&\sqrt{\frac{\Omega_a}{2\mu\omega}}\;p_1
+\sqrt{\frac{\mu\omega\Omega_a}{2}}\;x_2, \quad
P_b=\sqrt{\frac{\Omega_b}{2\mu\omega}}\;p_1
-\sqrt{\frac{\mu\omega\Omega_b}{2}}\;x_2,
\end{eqnarray}
where
\begin{equation}
\label{Eq:Om}\Omega_a=\omega(1+L_0),\quad \Omega_b=\omega(1-L_0).
\end{equation}
In the above $X_{\alpha}$ and $P_{\alpha}$ satisfy
$X_{\alpha}=X_{\alpha}^\dagger,$ $P_{\alpha}=P_{\alpha}^\dagger,$
$\left[X_{\alpha},X_{\beta}\right]=
\left[P_{\alpha},P_{\beta}\right]=0$ and
$\left[X_{\alpha},P_{\beta}\right]=i\delta_{{\alpha}{\beta}}.$
Then we define following annihilation-creation operators
$A_{\alpha}, A_{\alpha}^{\dagger}$
\begin{eqnarray}
\label{Eq:AA}
A_{\alpha}=i\sqrt{\frac{1}{2\Omega_{\alpha}}}\;P_{\alpha}
+\sqrt{\frac{\Omega_{\alpha}}{2}}\;X_{\alpha}, \quad
A_{\alpha}^{\dagger}=-i\sqrt{\frac{1}{2\Omega_{\alpha}}}\;P_{\alpha}
+\sqrt{\frac{\Omega_{\alpha}}{2}}\;X_{\alpha},\; (\alpha=a,b).
\end{eqnarray}
Here $A_{\alpha}$ and $A_{\alpha}^{\dagger}$ satisfy
$\left[A_{\alpha},A_{\beta}\right]=
\left[A_{\alpha}^{\dagger},A_{\beta}^{\dagger}\right]=0$, and
$\left[A_{\alpha},A_{\beta}^{\dagger}\right]=\delta_{{\alpha}{\beta}}.$
The eigenvalues $n_{\alpha}$ of the number operator
$N_{\alpha}=A_{\alpha}^{\dagger}A_{\alpha}$ are
$n_{\alpha}=0,1,2,\cdots.$ The angular momentum (\ref{Eq:L2}) is
rewritten in the form of two uncoupled modes of frequencies
$\Omega_a$ and $\Omega_b:$
\begin{equation}
\label{Eq:L3} \hat L =\frac{\Omega_b}{\omega}A_b^\dagger
A_b-\frac{\Omega_a}{\omega}A_a^\dagger A_a-L_0.
\end{equation}
Where the zero point angular momentum $L_0$ appears again. The
spectrum of $\hat L$ is
\begin{equation}
\label{Eq:l0} l_{(n_a,n_b)}=\frac{\Omega_b}{\omega}n_b
-\frac{\Omega_a}{\omega}n_a-L_0=n_b-n_a-(n_a+n_b+1)L_0.
\end{equation}
$l_{(n_a,n_b)}$ takes fractional values. We notice that, unlike
the case in commutative space, the absolute value of the lowest
angular momentum is not zero, it is
$|L_0|=\xi^{-2}\mu\omega\Lambda_{NC}^{-2}|d|.$ For the case
$n_a=n_b=n$ the eigenvalues are proportional to $L_0$
\begin{equation}
\label{Eq:l1} l_{(n,n)}=-(2n+1)L_0.
\end{equation}
From Eq.~(\ref{Eq:l0}) the intervals of the spectrum are
$\Delta l_{(\Delta n_a,\Delta
n_b)}=l_{(n_a^{\prime},n_b^{\prime})}-l_{(n_a,n_b)}=\Delta n_b-\Delta
n_a-(\Delta n_a+\Delta n_b)L_0$
where $\Delta
n_a=n_a^{\prime}-n_a$ and $\Delta n_b=n_b^{\prime}-n_b.$ There are
different intervals in the spectrum. For the case $\Delta
n_a=\Delta n_b=\Delta n$ we have fractional intervals
$\Delta l_{(\Delta n,\Delta n)}=-2\Delta n L_0.$
For the case $\Delta n_b=-\Delta n_a=\Delta n$
we have integer intervals
$\Delta l_{(-\Delta n, \Delta n)}=2\Delta n.$
We notice that the zero-point value $L_0,$ the spectrum
$l_{(n,n)}$ and the intervals $\Delta l_{(\Delta n,\Delta n)}$
take fractional values which are parameter $d-$ dependent. Such
fractional feature of the orbital angular momentum represents the
effects of noncommutative space.

In the limit $d\to 0$ we have $L_0\to 0,$ $\omega_a,\omega_b \to
\omega.$ Thus the spectrum $l_{(n_a,n_b)}\to
\omega\left(n_b-n_a\right)$ which recovers the results in
commutative space.

For consistent check, we use Eqs.~(\ref{Eq:hat-a-a1}) to
investigate the perturbation expansion of the first term in
Eq.~(\ref{Eq:L1}). We obtain
\begin{equation*}
-i\left(\hat a_1^\dagger\hat a_2 -\hat a_2^\dagger\hat
a_1\right)=-i\left(a_1^\dagger a_2 -a_2^\dagger
a_1\right)-\xi^{-2}\mu\omega\Lambda_{NC}^{-2} d
\left(a_1^\dagger a_1 +a_2^\dagger
a_2\right). \nonumber
\end{equation*}
The operators $a_1$ and $a_2$ are related to the operators $A_a$
and $A_b$ by the following equations
\begin{equation}
\label{Eq:aA} A_a=\frac{1}{\sqrt 2}\left(a_1+i a_2\right),\quad
A_b=\frac{1}{\sqrt 2}\left(a_1-i a_2\right).
\end{equation}
It follows that $-i\left(\hat
a_1^\dagger\hat a_2 -\hat a_2^\dagger\hat
a_1\right)=\frac{\Omega_b}{\omega}A_b^\dagger
A_b-\frac{\Omega_a}{\omega}A_a^\dagger A_a.$ Thus
Eq.~(\ref{Eq:L1}) equals Eq.~(\ref{Eq:L3}).

To conclude this paper we investigate the perturbation treatment of
the harmonic oscillator. Using Eq.~(\ref{Eq:hat-x-x}),
the perturbation expansion of Eq.~(\ref{Eq:H})
is
\begin{eqnarray}
\label{Eq:H2} \hat H(\hat x,\hat p)&=& \frac{1}{2\mu}p_i p_i
+\frac{1}{2}\mu\omega^2 x_i x_i -\xi^{-2}
\mu\omega^2\Lambda_{NC}^{-2}
d\epsilon_{ij}x_i p_j.
\end{eqnarray}
In the above the Chern-Simons term
$-\xi^{-2}\mu\omega^2\Lambda_{NC}^{-2} d\epsilon_{ij}x_i p_j$
represents the effect of noncommutative space. Because of the
scaling factor $\xi$ in Eqs.~(\ref{Eq:hat-x-x}) the corrections
from the kinetic energy term and the potential term cancel each
other, thus in Eq.~(\ref{Eq:H2}) there are no corrections to the
mass and frequency. These are different from the results obtained
in \cite{MM} where the effective mass and frequency have
$d-$dependent corrections.

Eq.~(\ref{Eq:H2}) is exactly solvable. By the same procedure of
obtaining (\ref{Eq:L3}) we rewrite the Hamiltonian (\ref{Eq:H2}) as
\begin{equation}
\label{Eq:H+H} H=H_a+H_b, \quad H_{\alpha}=
\Omega_{\alpha}\left(A_{\alpha}^\dagger
A_{\alpha}+\frac{1}{2}\right),\quad (\alpha=a,b).
\end{equation}
Because of $\Omega_a\ne \Omega_b$ the eigenvalues of $H$ are
non-degenerate.

In summary, in noncommutative space the maintenance of
Bose-Einstein statistics at the deformed level of $\hat a_i$ and
$\hat a_i^\dagger$ requires {\it simultaneously} space-space
noncommutativity and momentum-momentum noncommutativity. The
commutation relations (\ref{Eq:[a,a+]1}) and (\ref{Eq:[a,a+]2}),
and the perturbation expansions (\ref{Eq:hat-x-x}) and
(\ref{Eq:hat-a-a1}) are general, they provide a consistent
framework for the further development of quantum theories in
noncommutative space. Further exploration of the influence of the
new deformed commutation relations (\ref{Eq:[a,a+]2}) on dynamics
is interesting.

\vspace{0.4cm}

This work has been supported by the Deutsche
Forschungsgemeinschaft (Germany). The author would like to thank
W. R\"uhl for stimulating discussions.
 His work has also been supported by the National Natural Science
Foundation of China under the grant number 10074014 and by the Shanghai
Education Development Foundation.



\end{document}